\documentclass[onecolumn,showpacs,preprintnumbers,amsmath,amssymb]{revtex4}

\usepackage{epsf}
\usepackage{graphicx}  
\usepackage{dcolumn}   
\usepackage{bm}        
\usepackage{subfigure}

\begin{document}

\title{$\Lambda(t)$CDM Model as a Unified Origin of Holographic and Agegraphic Dark Energy Models}

\author{Yun Chen}
 \email{chenyun@mail.bnu.edu.cn}
\affiliation{ Department of Astronomy, Beijing Normal University,
Beijing 100875, China}

\author{Zong-Hong Zhu}
\email{zhuzh@bnu.edu.cn}
\affiliation{ Department of Astronomy, Beijing Normal University, Beijing 100875, China}

\author{Lixin Xu}
 \email{lxxu@dlut.edu.cn}
\affiliation{School of Physics and Optoelectronic Technology, Dalian
University of Technology, Dalian, 116024, China}

\author{J.S. Alcaniz}
\email{alcaniz@on.br}
 \affiliation{Departmento de Astronomia,
              Observat\'orio Nacional, 20921-400, Rio de Janeiro -- RJ,
              Brasil}

\date{ \today}


\begin{abstract}

 Motivated by the fact that any nonzero $\Lambda$ can
introduce a length scale or a time scale into Einstein's theory,
$r_{\Lambda}=ct_{\Lambda}=\sqrt{3/|\Lambda|}$. Conversely, any
cosmological length scale or time scale can introduce a $\Lambda
(t)$, $\Lambda(t)=3/r^2_{\Lambda}(t)=3/(c^2t^2_{\Lambda}(t))$. In
this letter, we investigate the time varying $\Lambda(t)$
corresponding to the length scales, including the Hubble horizon,
the particle horizon and the future event horizon, and the time
scales, including the age of the universe and the conformal time. It
is found out that, in this scenario, the $\Lambda(t)$CDM model can
be taken as the unified origin of the holographic and agegraphic
dark energy models with interaction between the matter and the dark
energy, where the interacting term is determined by
$Q=-\dot{\rho}_{\Lambda}$. We place observational constraints on the
$\Lambda(t)$CDM models originating from different cosmological
length scales and time scales with the recently compiled ``Union2
compilation'' which consists of 557 Type Ia supernovae (SNIa)
covering a redshift range $0.015\leq z \leq 1.4$. In conclusion, an
accelerating expansion universe can be derived in the cases taking
the Hubble horizon, the future event horizon, the age of the
universe and the conformal time as the length scale or the time
scale.

\end{abstract}


\maketitle
\section{Introduction}
 The accelerating expansion of the universe is one of the most important issues of modern cosmology,
 which has been discovered and verified by supernova \cite{SN}, CMB \cite{CMB} and
 BAO \cite{BAO} observations (see also recent reviews \cite{Frieman08,Tsujikawa2010}). After
the discovery of this scenario, a great variety of attempts have
been done to explain this acceleration (see the reviews
\cite{Rev,Tsujikawa2010}). Among these alternatives, $\Lambda$CDM is
the simplest and most nature one, which fits the observational data
best. In this scenario, the dark energy is associated to the energy
density of the quantum vacuum. Briefly put, the dark energy is the
energy stored on the true vacuum state of all existing fields in the
universe, i.e. $\rho_\Lambda=\Lambda/8\pi G$, where $\Lambda$ is the
cosmological constant. However, it is embarrassed by cosmological
constant problems, namely the fine-tuning problem and the
coincidence problem \cite{ccp}. Several possible approaches have be
adopted to explain or alleviate the cosmological constant problems
\cite{accp}, though there is no convincing fundamental theory for
why vacuum energy dominance happened only recently. The possibility
that if $\Lambda$ is not a real constant but is time variable or
dynamical, the cosmological problems may be alleviated or removed,
was considered many years ago \cite{Peebles1988, wDE-1}. This kind
of models can be called as $\Lambda(t)$CDM models. The traditional
approach for $\Lambda(t)$CDM models was first to specify a
phenomenological time-dependence form for $\Lambda(t)$ and then
establish a cosmological scenario. There are a lot of proposals of
the phenomenological forms for $\Lambda(t)$ in the literature
\cite{rhoP}, such as $\rho_{\Lambda}=\sigma H$ and
$\rho_{\Lambda}=n_1H+n_2H^2$. In our scenario, $\Lambda(t)$ is
determined based on its relation with the cosmological length scale
or time scale.

As it is known, in the de Sitter universe, any nonzero $\Lambda$ can
introduce a length scale $r_{\Lambda}$ and time scale $t_{\Lambda}$,
as the form
\begin{equation}
\label{eq:rt} r_{\Lambda}=ct_{\Lambda}=\sqrt{3/|\Lambda|},
\end{equation}
into Einstein's theory \cite{Bousso08}, where $c$ is the speed of
light and throughout this letter we use the unit $c=1$. Conversely,
a cosmological length scale and time scale may introduce a $\Lambda
(t)$ into Einstein's theory
\begin{equation}
\label{eq:lambda}
\Lambda(t)=\frac{3}{r^2_{\Lambda}(t)}=\frac{3}{t^2_{\Lambda}(t)}.
\end{equation}
Obviously, when the length scale or time scale is time variable, a
time varying $\Lambda$ can be obtained. The key problem is how to
choose a proper cosmological length scale or time scale to obtain a
tiny $\Lambda$. However, we have not acquired the first physical
principle to determine the length or time scale recently.
Nevertheless, one can immediately relate these length or time scales
to the biggest natural length scales, including the Hubble horizon,
the particle horizon and the future event horizon, and natural time
scales, including the age of the universe and the conformal time.
The cosmological constant derived from the length scale can be
called as \textit{horizon cosmological constant} \cite{xu09a}, and
the one derived from the time scale can be called as \textit{age
cosmological constant} \cite{xu09b}.

Once mentioning the length scale and time scale, one may easily
associate them with the holographic and agegraphic dark energy.
Based on the holographic principle, Cohen et al. \cite{Cohen99}
suggested a relation between the IR cut-off and the UV cut-off in
the quantum field theory. For a system with size $L$ representing
the IR cut-off and UV cut-off $\Lambda$ relating to the quantum
zero-point energy, without decaying into a black hole of the same
size, it is required that the total energy in a region of size $L$
should not exceed the mass of a black hole of the same size, namely
$L^3\rho_{\Lambda}\leq LM_P^2$. When this inequality is reduced to
an equality
\begin{equation}
\label{eq:rhol1} \rho_{\Lambda}=3c^2M^2_{P}L^{-2},
\end{equation}
$L$ takes the maximum value, where $M_P\equiv 1/\sqrt{8\pi G}$ is
the reduced Planck mass and $3c^2$ is a numerical factor. Recently,
three length scales of the universe: the Hubble horizon, the
particle horizon and the future event horizon, have been taken as
the role of IR cut-off. However, when the Hubble horizon and the
particle horizon are chosen as the IR cut-off, non-accelerated
expansion universe can be achieved. Li \cite{Li04} proposed to take
the future event horizon as the IR cut-off leading to the
holographic dark energy model which can derive an accelerated
expansion universe. Also, this model has been constrained with
different observational data and is consistent with the data
\cite{holo}. However, it is embarrassed on the fundamental level due
to its assumption that the current properties of the dark energy are
determined by the future evolution of the universe, which is the
so-called causality problem \cite{Cai07}. What's more, it has been
pointed out that this model can be inconsistent with the age of the
universe \cite{WZ07}. Recently, based on the K\'arolyh\'azy
uncertainty relation $\delta t = \beta t^{2/3}_p t^{1/3}$ together
with the time-energy uncertainty relation, also called the
Heisenberg uncertainty relation $E_{\delta t^3} \sim t^{-1}$, one
can estimate the quantum energy density of the metric fluctuations
of Minkowski space-time that is $\rho_{\Lambda}=3n^2M^2_P/t^2$,
where $3n^2$ is a numerical factor representing some uncertainties.
In the above context, $\beta$ is a numerical factor of order one,
$t_p$ is the reduced Planck time, and throughout this letter, the
units $c=\hbar
 =k_b=1$ are adopted, so that one has $l_p=t_p=1/m_p$ with $l_p$ and $m_p$ being the
 reduced Planck length and mass, respectively. Recently,
two time scales of the universe: the age of the universe and the
conformal time, have been taken as the role of $t$, corresponding to
the agegraphic dark energy \cite{Cai07} and the new agegraphic dark
energy \cite{WC08a}. Both of the models can derive an accelerated
expansion universe, and is consistent with the recent observational
data \cite{agegra}. They two also can resolve the causality problem,
however, it has been presented that the agegraphic dark energy model
is classically unstable, and the new agegraphic dark energy model is
no better than the holographic dark energy model for the description
of the dark energy-dominated universe \cite{agegraP}. After a brief
introduction of the holographic and agegraphic dark energy, one may
be more interested in that whether there are some relationships
between them and the $\Lambda(t)$CDM models investigated in this
letter. The corresponding problems will be discussed in Sections III
and IV.

The letter is organized as follows. In Section II, the basic
equations of the time variable cosmological constant models are
given. In Section III, three length scales and two time scales will
be considered to derive the time variable cosmological constant
models, and constraints from the recent observational data are
illustrated. Furthermore, the relationships between the time
variable cosmological constant models and the holographic dark
energy and agegraphic dark energy are investigated. Finally, we
provide our conclusion and discussion in Section IV.

\section{$\Lambda(t)$CDM cosmology: basic equations}
Considering the Einstein field equation
\begin{equation}
\label{eq:EE} R_{\mu\nu}-\frac{1}{2}Rg_{\mu\nu}+\Lambda
g_{\mu\nu}=8\pi G T_{\mu\nu},
\end{equation}
where $T_{\mu\nu}$ is the energy-momentum tensor of matter and
radiation, one sees by Bianchi identities that when the
energy-momentum tensor is conserved, namely
$\nabla^{\mu}T_{\mu\nu}=0$, it follows necessarily that $\Lambda$ is
a constant. To accommodate the running of $\Lambda$ with the cosmic
time, namely $\Lambda=\Lambda(t)$, the most obvious way is to shift
$\Lambda (t)$ to the right-hand side of Eq. (\ref{eq:EE}) and to
take $\tilde{T}_{\mu\nu}=T_{\mu\nu}-\frac{\Lambda(t)}{8\pi
G}g_{\mu\nu}$ as the total energy-momentum tensor. By requiring the
local energy-momentum conservation law,
$\nabla^{\mu}\tilde{T}_{\mu\nu}=0$, one yields
\begin{equation}
\label{eq:Tconservation} \dot{\rho}_{T}+3H(\rho_T+p_T)=0,
\end{equation}
with $H=\dot{a}/a$ being the Hubble parameter, where an over dot
means taking derivative with respect to the cosmic time $t$. In a
spacially flat FRW universe ignoring the radiation, with
$\rho_T=\rho_m+\rho_\Lambda$ and $p_T=p_m+p_\Lambda$, the Friedmann
equation can be written as
\begin{equation}
\label{eq:FE}
H^2=\frac{1}{3M^2_P}\left(\rho_{m}+\rho_{\Lambda}\right).
\end{equation}
With the definitions of the dimensionless density parameters
$\Omega_m=\rho_m/3M^2_pH^2$ and
$\Omega_{\Lambda}=\rho_{\Lambda}/3M^2_pH^2$, one can reduce Eq.
(\ref{eq:FE}) to
\begin{equation}
\label{eq:FE1} \Omega_{\Lambda}+\Omega_m=1.
\end{equation}
In this situation, Eq. (\ref{eq:Tconservation}) is reduced to
\begin{equation}
\label{eq:conservation}
\dot{\rho}_{\Lambda}+\dot{\rho}_{m}+3H\left(1+w_{m}\right)\rho_{m}+3H(1+w_{\Lambda})\rho_{\Lambda}=0.
\end{equation}
Throughout, the subscript ``$m$'' denotes the corresponding quantity
of the matter which contains the baryonic and dark matter, as well
as, ``$\Lambda$'' represents the corresponding quantity of the
vacuum energy and ``T'' indicates the corresponding quantity of
total components. $w_i=p_i/\rho_i$ is the equation of state (EoS) of
the \textit{i}th-component (here $i= m, \Lambda$), where $\rho_i$
and $p_i$ are density and pressure of corresponding component, with
$\rho_{\Lambda}=M^2_{P}\Lambda(t)$.

Assuming the vacuum energy and matter exchange energy through
interaction term $Q$,  then Eq. (\ref{eq:conservation}) can be
rewritten as
\begin{equation}
\label{eq:conservDM} \dot{\rho}_{m}+3H(1+w_m)\rho_{m}=Q,
\end{equation}
and
\begin{equation}
\label{eq:conservDE}
\dot{\rho}_{\Lambda}+3H(1+w_{\Lambda})\rho_{\Lambda}=-Q.
\end{equation}
Rewriting Eqs. (\ref{eq:conservDM}) and (\ref{eq:conservDE}) as
\begin{equation}
\label{eq:conservDM1} \dot{\rho}_{m}+3H(1+w_m^{eff})\rho_{m}=0
\end{equation}
and
\begin{equation}
\label{eq:conservDE1}
\dot{\rho}_{\Lambda}+3H(1+w_{\Lambda}^{eff})\rho_{\Lambda}=0,
\end{equation}
one can define the effective EoS for the matter and the vacuum
energy as
\begin{equation}
\label{eq:wmeff} w_m^{eff}=w_m-\frac{Q}{3H\rho_m},
\end{equation}
and
\begin{equation}
\label{eq:woleff}
w_{\Lambda}^{eff}=w_{\Lambda}+\frac{Q}{3H\rho_{\Lambda}}.
\end{equation}
For the pressureless matter, one reads $p_m=0$ and $w_m=0$. Within
this framework, it is interesting to mention that the EoS of the
vacuum energy takes the usual form
$w_{\Lambda}(t)=p_{\Lambda}(t)/\rho_{\Lambda}(t)=-1$ \cite{wDE-1}.
It shows that the value of this EoS does not depend on whether
$\Lambda$ is strictly constant or time variable. Furthermore, Eqs.
(\ref{eq:conservDM}) and (\ref{eq:conservDE}) are reduced as
\begin{equation}
\label{eq:conservDE2} \dot{\rho}_{\Lambda}=-Q
\end{equation}
and
\begin{equation}
\label{eq:conservDM2}
\dot{\rho}_{m}+3H\rho_{m}=-\dot{\rho}_{\Lambda},
\end{equation}
which shows that the matter and the vacuum energy are not
independently conserved, with the decaying vacuum density
$\rho_{\Lambda}$ playing the role of a source of matter production.
In this regard, the $\Lambda (t)$CDM model also can be called as the
decaying vacuum model. Jointing Eqs.  (\ref{eq:wmeff}),
(\ref{eq:woleff}) and (\ref{eq:conservDE2}), one obtains
\begin{eqnarray}
 w_{\Lambda}^{eff}& = &-1-\frac{\dot{\rho}_{\Lambda}}{3H\rho_{\Lambda}}\nonumber\\
 & = &-1-\frac{1} {3}\frac{d\ln \rho_{\Lambda}}{d\ln a}\label{eq:woleff1}
\end{eqnarray}
and
\begin{eqnarray}
w_{m}^{eff}& = &-(w_{\Lambda}^{eff}+1)\frac{\rho_{\Lambda}}{\rho_m}\nonumber\\
 & = &-(w_{\Lambda}^{eff}+1)\frac{\Omega_{\Lambda}}{1-\Omega_{\Lambda}},\label{eq:wmeff1}
\end{eqnarray}
Using Eqs. (\ref{eq:FE}) and (\ref{eq:conservDM1}), one gets the
following solution
\begin{equation}
\label{eq:E2} E^2(z)=\frac{(1-\Omega_{\Lambda
0})\exp\{3\int_0^z\frac{dz'}{1+z'}[1+w_m^{eff}(z')]\}}{1-\Omega_{\Lambda}},
\end{equation}
where $E=H/H_0$ is the dimensionless Hubble parameter. Throughout
the subscript ``$0$'' denotes the value of a quantity today.
Moreover, substituting Eq. (\ref{eq:FE}) into Eq.
(\ref{eq:conservation}), one yields
\begin{equation}
\label{eq:H1}
 \frac{d\ln H}{d\ln a}+\frac{3}{2}(1-\Omega_{\Lambda})=0.
\end{equation}

According to the definitions of the decelerating parameter $q\equiv
-(\ddot{a}a)/{\dot{a}}^2$ and the Hubble parameter $H\equiv
\dot{a}/a$, one can obtain
\begin{equation}
\label{eq:q1} q=(-\frac{\ddot{a}}{a})/H^2=-1-\frac{d\ln H}{d\ln a}.
\end{equation}
Furthermore, from Eqs.(\ref{eq:H1}) and (\ref{eq:q1}), one reads
\begin{equation}
\label{eq:q2} q=\frac{1}{2}-\frac{3\Omega_{\Lambda}}{2},
\end{equation}
which shows that to ensure the current accelerating expansion of the
universe, $\Omega_{\Lambda 0}>1/3$ is required. With
Eq.(\ref{eq:woleff1}) and the definition of the dimensionless energy
density $\Omega_{\Lambda}=\rho_{\Lambda}/3M^2_pH^2$, the effective
EoS of the vacuum energy is given by
\begin{equation}
\label{eq:woleff2} w_{\Lambda}^{eff}=-1-\frac{1}{3}(\frac{d\ln
\Omega_{\Lambda}}{d\ln a}+2\frac{d\ln H}{d\ln a}).
\end{equation}
Moreover, with Eqs.(\ref{eq:H1}) and (\ref{eq:woleff2}), one obtains
\begin{equation}
\label{eq:woleff3}
w_{\Lambda}^{eff}=-\Omega_{\Lambda}-\frac{1}{3}\frac{d\ln
\Omega_{\Lambda}}{d\ln a}.
\end{equation}
Substituting Eq. (\ref{eq:woleff3}) into Eq. (\ref{eq:wmeff1}), one
gets
\begin{equation}
\label{eq:wmeff2}
w_{m}^{eff}=\frac{\Omega_{\Lambda}}{1-\Omega_{\Lambda}}(\Omega_{\Lambda}+\frac{1}{3}\frac{d\ln
\Omega_{\Lambda}}{d\ln a}-1).
\end{equation}

\section{Constraints from the recent observational data}

\subsection{Horizon cosmological constant}

\subsubsection{Hubble horizon as a cosmological length scale}
When Hubble horizon $H^{-1}$ is chosen, one obtains
\begin{equation}
\Lambda(t)=3c^2H^{2}(t)
\end{equation}
where $c$ is a constant. As it is known, our universe is filled with
the matter and dark energy which deviates from a de Sitter universe.
Just to describe this gap, the constant $c$ was introduced in
\cite{xu09a}. Naturally, when $c=1$, the de Sitter universe will be
recovered. Now, according to the definition of the energy density
$\rho_{\Lambda}=M^2_P\Lambda (t)$, the corresponding vacuum energy
density can be written as
\begin{equation}
\label{eq:rhol}\rho_{\Lambda}=3c^2M^2_{P}H^2
\end{equation}
which takes the same form as the holographic dark energy with the
Hubble horizon based on holographic principle\cite{Li04}. With this
vacuum energy, the Friedmann equation (\ref{eq:FE}) can be rewritten
as
\begin{equation}
\label{eq:rhom}\rho_{m}=3(1-c^2)M^2_PH^2.
\end{equation}
Requiring a positive value for the matter energy density $\rho_m$,
the condition
\begin{equation}
\label{eq:c1}c^2<1,
\end{equation}
must be satisfied. According to the definition of the dimensionless
energy density, one has
\begin{eqnarray}
\label{eq:olhh} \Omega_{\Lambda}=\frac{\rho_{\Lambda}}{\rho_c}
=\frac{3c^2M^2_pH^2}{3M^2_pH^2}=c^2.
\end{eqnarray}
In this case, from Eqs.(\ref{eq:olhh}) and (\ref{eq:q2}), the
decelerating parameter becomes
\begin{equation}
\label{eq:qhh} q=\frac{1}{2}-\frac{3}{2}c^2.
\end{equation}
To obtain a current accelerating expansion universe, i.e. $q<0$, and
to ensure a positive matter energy density, the condition
\begin{equation}
\label{eq:c2}1/3<c^2<1
\end{equation}
is necessary. Furthermore, from Eqs.(\ref{eq:olhh}) and
(\ref{eq:woleff3}), the effective EoS of vacuum energy density is
\begin{equation}
\label{eq:whh} w_{\Lambda}^{eff}=-c^2.
\end{equation}
With the condition Eq. (\ref{eq:c2}), one can see that
$-1<w_{\Lambda}^{eff}<-1/3$, which means that a quintessence-like
dark energy is obtained. Substituting Eq. (\ref{eq:olhh}) into Eq.
(\ref{eq:wmeff2}), one read
\begin{equation}
\label{eq:wmhh} w_{m}^{eff}=-c^2.
\end{equation}
With Eqs. (\ref{eq:E2}), (\ref{eq:olhh}) and (\ref{eq:wmhh}), the
Friedmann equation of this model is reduced to
\begin{equation}
\label{eq:E2hh} E^2(z;c)=(1+z)^{3(1-c^2)},
\end{equation}
which shows that there is only one model-dependent parameter $c$.

\begin{figure}[htp]
$\begin{array}{cc}
\subfigure[]{\includegraphics[width=1.1\textwidth]{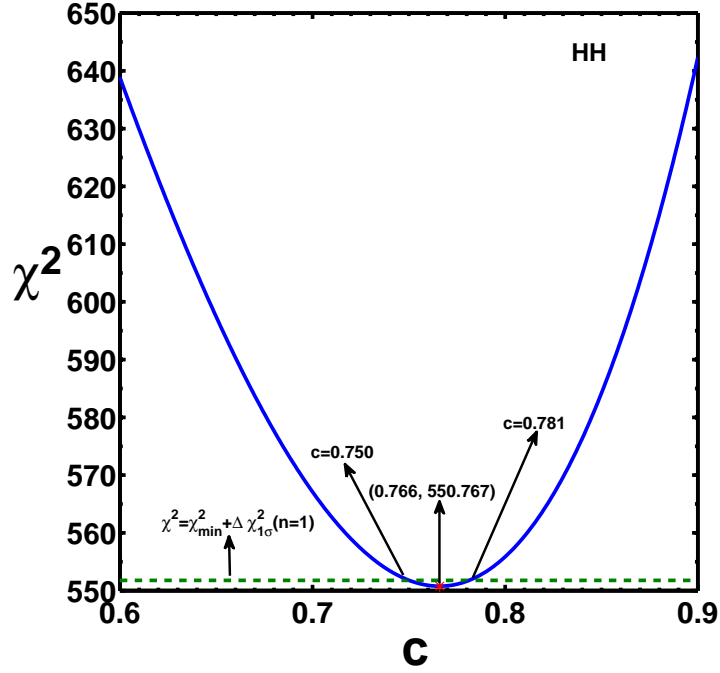}

\label{fig:chi2HH}}\\ \subfigure[]{
\includegraphics[width=1.1\textwidth]{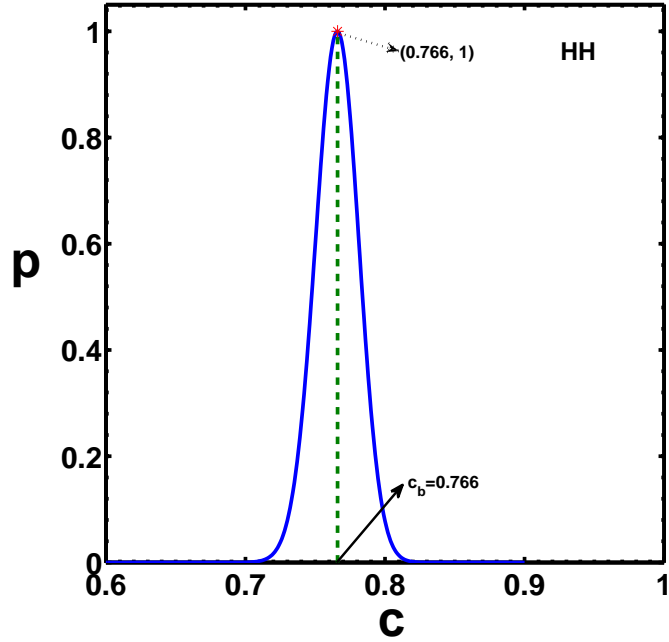} \label{fig:likelihood}}
\end{array}$
\caption{Hubble horizon is taken as a cosmological length scale (HH
for short). The best-fit value for parameter $c$ is $c_b=0.766$ with
$\chi^2_{min}=550.767$. The confidence range is $0.750\leq
 c\leq 0.781$, i.e. $c=0.766^{+0.015}_{-0.016}$ with 68.3\% confidence level.
  Fig. \ref{fig:chi2HH} shows the evolution of $\chi^2$ with respect to the
  parameter $c$. The green dashed line denotes
  $\chi^2=\chi^2_{min}+\Delta\chi^2_{1\sigma}(n=1)$. The red dot
  indicates the best-fit pair $(c, \chi_min)=(0.766, 550.767)$.
  Fig. \ref{fig:likelihood} displays the probability distribution of $c$, where the probability is defined as
  $p=\exp(-\chi^2/2)/\exp(-\chi_{min}^2/2)$. The green dashed line denotes
  $c=c_b$.} \label{fig:HH}
\end{figure}

We perform the $\chi^2$ statistics to constrain the parameter $c$ in
this model with the recently compiled ``Union2 compilation'' which
consists of 557 Type Ia supernovae (SNIa) covering a redshift range
$0.015\leq z \leq 1.4$ \cite{Amanullah2010}. The best-fit value for
parameter $c$ is $c_b=0.766$ with $\chi^2_{min}=550.767$. The
confidence range is $0.750\leq
 c\leq 0.781$, i.e. $c=0.766^{+0.015}_{-0.016}$ with 68.3\% confidence
 level. Furthermore, one can work out $\Omega_{m0}=0.413^{+0.025}_{-0.023}$
 and $\Omega_{\Lambda 0}=0.587^{+0.023}_{-0.025}$ with 68.3\% confidence
 level. Fig. \ref{fig:chi2HH} shows the evolution of $\chi^2$ with respect
to the parameter $c$. Fig. \ref{fig:likelihood} displays the
probability distribution of $c$, where the probability is defined as
  $p=\exp(-\chi^2/2)/\exp(-\chi_{min}^2/2)$.

\subsubsection{Particle horizon as a cosmological length scale}
The particle horizon is defined as
\begin{equation}
\label{eq:Rp}R_{p}=a(t)\int^{t}_{0}\frac{dt'}{a}=a\int^{a}_{0}\frac{da'}{Ha'^{2}}
\end{equation}
which is the length scale a particle can pass from the beginning of
the universe $t=0$. When the particle horizon $R_{p}$ is chosen, one
gets
\begin{equation}
\label{eq:ltRp}\Lambda(t)=3c^2/R_{p}^2
\end{equation}
In this case, the vacuum energy density is given as
\begin{equation}
\label{eq:rhoRp}\rho_{\Lambda}=3c^2M^2_P/R^2_p,
\end{equation}
which takes the same form as the holographic dark energy with the
particle horizon\cite{Li04}, where $c$ is taken to fill the
deviation from the de Sitter universe. Combing Eqs.(\ref{eq:Rp}),
(\ref{eq:rhoRp}) and the definition of the dimensionless energy
density $\Omega_\Lambda$, one has
\begin{equation}
\label{eq:Rp2} \int^{a}_{0}\frac{d\ln
a'}{Ha'}=\frac{c}{aH}\sqrt{\frac{1}{\Omega_{\Lambda}}}.
\end{equation}
Taking the derivative with respect $\ln a$ from the both sides of
the above Eq.(\ref{eq:Rp2}), one has the differential equation
\begin{equation}
\label{eq:FriedmannRp} \frac{d\ln H}{d\ln
a}+\frac{1}{2\Omega_\Lambda}\frac{d\Omega_\Lambda}{d\ln
a}=-\frac{\sqrt{\Omega_\Lambda}}{c}-1
\end{equation}
Substituting Eq.(\ref{eq:H1}) into the above
Eq.(\ref{eq:FriedmannRp}), one obtains the differential equation of
$\Omega_\Lambda$
\begin{equation}
\label{eq:olRp}\frac{d\Omega_\Lambda}{dz}=-\Omega_\Lambda
(1-3\Omega_\Lambda-\frac{2}{c}\sqrt{\Omega_\Lambda})(1+z)^{-1}.
\end{equation}
Once the values of $c$ and $\Omega_{\Lambda0}$ are fixed, the
evolution of $\Omega_{\Lambda}$ as a function of the redshift $z$
can be obtained. From Eqs.(\ref{eq:woleff3}) and (\ref{eq:olRp}), it
is easy to get
\begin{equation}
\label{eq:wRp}
w_{\Lambda}^{eff}=-\frac{1}{3}+\frac{2}{3c}\sqrt{\Omega_\Lambda}.
\end{equation}
Obviously, $w_{\Lambda}^{eff}>-1/3$, hereby, the particle horizon
could not describe the accelerating phase.

\subsubsection{Future event horizon as a cosmological length scale}
The future event horizon is defined as
\begin{equation}
\label{eq:Re}
R_{e}=a\int^{\infty}_{t}\frac{dt'}{a}=a\int^{\infty}_{a}\frac{da'}{Ha'^{2}}
\end{equation}
which is the boundary of the volume a fixed observer may eventually
observe\cite{Li04}. Taking it as the role of cosmological length
scale, one has the vacuum energy density
\begin{equation}
\label{eq:rhoRe}\rho_{\Lambda}=3c^2M^2_P/R^2_e,
\end{equation}
which takes the same form as the holographic dark energy with the
future event horizon\cite{Li04}, where the constant $c$ is also
taken to fill the deviation from the de Sitter universe. Combining
Eq. (\ref{eq:Re}), Eq. (\ref{eq:rhoRe}) and the definition of the
dimensionless energy density $\Omega_{\Lambda}$, one has
\begin{equation}
\label{eq:Re2} \int^{\infty}_{a}\frac{d\ln
a'}{Ha'}=\frac{c}{aH}\sqrt{\frac{1}{\Omega_{\Lambda}}}.
\end{equation}
Taking the derivative with respect to $\ln a$ from the both sides of
the above equation (\ref{eq:Re2}), one has the differential equation
\begin{equation}
\label{eq:FriedmannRe}\frac{d\ln H}{d\ln
a}+\frac{1}{2}\frac{d\ln\Omega_{\Lambda}}{d\ln
a}=\frac{\sqrt{\Omega_{\Lambda}}}{c}-1.
\end{equation}
Substituting Eq. (\ref{eq:H1}) into the above differential equation,
one obtains
\begin{equation}
\label{eq:olRe} \frac{d\Omega_\Lambda}{dz}=-\Omega_\Lambda
(1-3\Omega_\Lambda+\frac{2}{c}\sqrt{\Omega_\Lambda})(1+z)^{-1}
\end{equation}
With Eqs.(\ref{eq:woleff3}) and (\ref{eq:olRe}), one reads
\begin{equation}
\label{eq:wRe}
w_{\Lambda}^{eff}=-\frac{1}{3}-\frac{2}{3c}\sqrt{\Omega_\Lambda}.
\end{equation}
Obviously, $w_{\Lambda}^{eff}<-1/3$, hence, the future event horizon
can lead to the accelerating phase. The decelerating parameter
$q(z)$ is expressed as Eq.(\ref{eq:q2}), where $\Omega_{\Lambda}$ is
determined by Eq. (\ref{eq:olRe}). Substituting Eq. (\ref{eq:olRe})
into Eq. (\ref{eq:wmeff2}), one reads
\begin{equation}
\label{eq:wmRe}
w_{m}^{eff}=-\frac{2}{3}(1-\frac{\sqrt{\Omega_{\Lambda}}}{c})\frac{\Omega_{\Lambda}}{1-\Omega_{\Lambda}}.
\end{equation}
 The Friedmann equation of this
model is determined by jointing Eqs. (\ref{eq:E2}), (\ref{eq:olRe})
and (\ref{eq:wmRe}).

\begin{figure}
 \centering
 \includegraphics[totalheight=3.5in, angle=0]{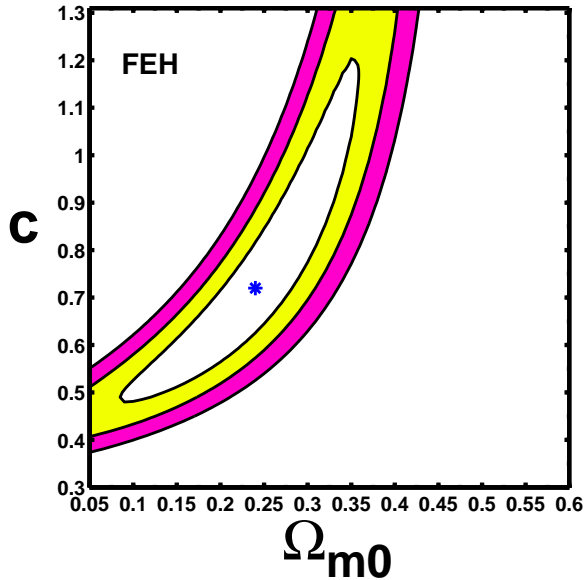}
 \caption{
 Future event horizon is taken as a
cosmological length scale (FEH for short). The contours correspond
to $68.3\%$, $95.4\%$ and $99.7\%$ confidence levels constrained
from the recent observational data in $(\Omega_{m0}, c)$ plane. The
blue dot marks the best-fit pair $(\Omega_{m0}, c)$. The best-fit
parameters in this case are found to be
$\Omega_{m0}=0.24^{+0.11}_{-0.16}$ and $c=0.72^{+0.49}_{-0.24}$
($68.3\%$ c.l.) with $\chi^2_{min}=544.367$.}
 \label{fig:FEH}
 \end{figure}

We perform the $\chi^2$ statistics to constrain the parameters
$(\Omega_{m0}, c)$ with the recently compiled ``Union2 compilation''
of SNeIa data. Figure \ref{fig:FEH} shows the probability contours
constrained from the observational data in $(\Omega_{m0} , c)$
plane. The best-fit parameters in this case are found to be
$\Omega_{m0}=0.24^{+0.11}_{-0.16}$ and $c=0.72^{+0.49}_{-0.24}$ for
$68.3\%$ confidence level with $\chi^2_{min}=544.367$.

\subsection{Age cosmological constant}

\subsubsection{Age of the universe as a cosmological time scale}
The age of the universe is defined as
\begin{equation}
\label{eq:tAU}t_{\Lambda}=\int^t_{0}dt'=\int^a_{0}\frac{da'}{a'H}.
\end{equation}
Taking it as the role of time scale, one obtains
\begin{equation}
\label{eq:ltAU} \Lambda (t)=3c^2/t_{\Lambda}^2,
\end{equation}
where $c$ is the model constant to fill the derivation from the de
Sitter universe\cite{xu09b}. In this case, one has the vacuum energy
density
\begin{equation}
\label{eq:rhoAU}\rho_{\Lambda}=3c^2M^2_P/t^2_{\Lambda},
\end{equation}
which takes the same form as the agegraphic dark energy\cite{Cai07}.
Combining Eq. (\ref{eq:tAU}), Eq. (\ref{eq:rhoAU}) and the
definition of the dimensionless energy density $\Omega_{\Lambda}$,
one has
\begin{equation}
\label{eq:tAU2}\int^{a}_{0}\frac{d\ln
a'}{H}=\frac{c}{H}\sqrt{\frac{1}{\Omega_{\Lambda}}}.
\end{equation}
Taking the derivative with respect to $\ln a$ from the both sides of
Eq.(\ref{eq:tAU2}), one has
\begin{equation}
\label{eq:FriedmannAU}\frac{d\ln H}{d\ln
a}+\frac{1}{2}\frac{d\ln\Omega_{\Lambda}}{d\ln
a}+\frac{\sqrt{\Omega_{\Lambda}}}{c}=0.
\end{equation}
Substituting Eq. (\ref{eq:H1}) into Eq.(\ref{eq:FriedmannAU}), one
gets
\begin{equation}
\label{eq:olAU} \frac{d\Omega_\Lambda}{dz}=-\Omega_\Lambda
(3-3\Omega_\Lambda-\frac{2}{c}\sqrt{\Omega_\Lambda})(1+z)^{-1}
\end{equation}
From Eqs.(\ref{eq:woleff3}) and (\ref{eq:olAU}), one has
\begin{equation}
\label{eq:wAU}
w_{\Lambda}^{eff}=-1+\frac{2}{3c}\sqrt{\Omega_\Lambda}.
\end{equation}
The accelerating universe requires $w_{\Lambda}^{eff}<-1/3$, that
is, $c>\sqrt{\Omega_\Lambda}$. With $0\leq\Omega_\Lambda\leq1$,
naturally, if $c>1$ the universe will eternally accelerate. For
$c<1$, the expansion would finally slow down in the future. With
Eqs. (\ref{eq:wmeff2}) and (\ref{eq:olAU}), one yields
\begin{equation}
\label{eq:wmAU}
w_{m}^{eff}=-\frac{2\sqrt{\Omega_\Lambda}}{3c}\frac{\Omega_\Lambda}{1-\Omega_\Lambda}.
\end{equation}
Jointing Eqs. (\ref{eq:E2}), (\ref{eq:olAU}) and (\ref{eq:wmAU}),
one can determine the Friedmann equation of this model.

\begin{figure}
 \centering
 \includegraphics[totalheight=3.5in, angle=0]{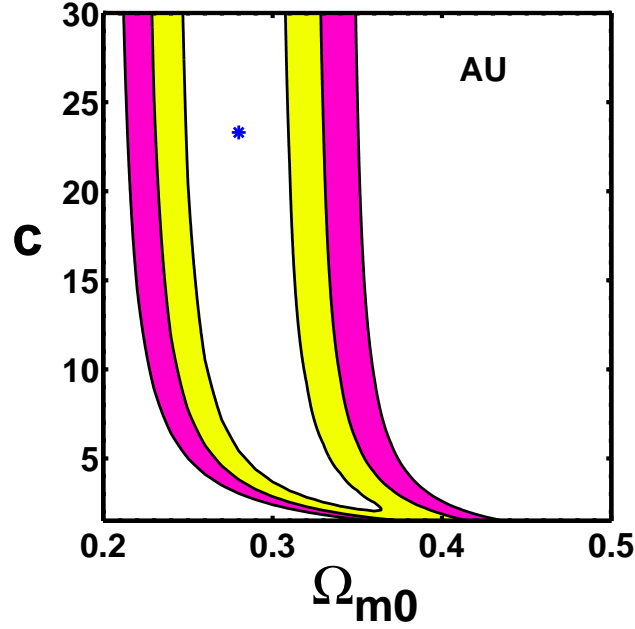}
 \caption{
Age of the universe is taken as a cosmological time scale (AU for
short). The contours correspond to $68.3\%$, $95.4\%$ and $99.7\%$
confidence levels constrained from the recent observational data.
The blue dot marks the best-fit pair $(\Omega_{m0}, c)$. The results
are $\Omega_{m0}=0.28^{+0.08}_{-0.07}$ and $c=23.3^{+null}_{-21.3}$
($68.3\%$ c.l.) with $\chi^{2}_{min}=544.25$, where ``null'' denotes
the absence of the upper limit for $c$.}
 \label{fig:AU}
 \end{figure}

Using the recently compiled ``Union2 compilation'' of SNeIa data, We
perform the $\chi^2$ statistics to constrain the parameters
$(\Omega_{m0}, c)$ in this model. The best-fit results are
$\Omega_{m0}=0.28^{+0.08}_{-0.07}$ and $c=23.3^{+null}_{-21.3}$ for
$68.3\%$ confidence level with $\chi^2_{min}=544.25$, where ``null''
denotes the absence of the upper limit for $c$. Figure \ref{fig:AU}
displays the probability contours constrained from the observational
data in $(\Omega_{m0} , c)$ plane. However the top sides of the the
three contours are not closed, which denotes that we cannot get the
upper limit for $c$ in $68.3\%$, $95.4\%$ and $99.7\%$ confidence
levels. As a general illustration, we work out
$\chi^2_{c=\infty}=544.42$ with the best-fit value
$\Omega_{m0}=0.28$. Obviously,
$\Delta\chi^2=\chi^2_{c=\infty}-\chi^2_{min}=0.17$ is smaller than
$\Delta \chi^2_{1\sigma}(n=2)=2.30$ where $n$ is the number of the
parameters in the model, which implies that $c=\infty$ is still
inside the $1\sigma$ contour. Basically, this problem originates
from the evolution of $\Omega_{\Lambda}$ with respect to redshift
$z$ with the variety of $c$, that will be further discussed in
Section IV.

\subsubsection{Conformal time as a cosmological time scale}
The conformal time is defined as
\begin{equation}
\label{eq:tCT}
\eta_{\Lambda}=\int^{t}_{0}\frac{dt'}{a}=\int^{a}_{0}\frac{da'}{a'^{2}H},
\end{equation}
which is the maximum comoving distance to a comoving observer's
particle horizon since $t=0$\cite{MS09}. In this case, one has
\begin{equation}
\rho_{\Lambda}=3c^2M^2_P/\eta^2_{\Lambda}\label{eq:rhoCT},
\end{equation}
which takes the same form as the new agegraphic dark
energy\cite{WC08a}. Combining Eq. (\ref{eq:tCT}), Eq.
(\ref{eq:rhoCT}) and the definition of the dimensionless energy
density $\Omega_{\Lambda}$, one has
\begin{equation}
\label{eq:tCT2}\int^{a}_{0}\frac{d\ln
a'}{a'H}=\frac{c}{H}\sqrt{\frac{1}{\Omega_{\Lambda}}}.
\end{equation}
Taking the derivative with respect to $\ln a$ from the both sides of
Eq.(\ref{eq:tCT2}), one gets
\begin{equation}
\label{eq:FriedmannCT}\frac{d\ln H}{d \ln
a}+\frac{1}{2}\frac{d\ln\Omega_{\Lambda}}{d \ln
a}+\frac{\sqrt{\Omega_{\Lambda}}}{ac}=0.
\end{equation}
With Eqs. (\ref{eq:H1}) and (\ref{eq:FriedmannCT}), one obtains
\begin{equation}
\label{eq:olCT} \frac{d\Omega_\Lambda}{dz}=-\Omega_\Lambda
[\frac{3(1-\Omega_\Lambda)}{1+z}-\frac{2}{c}\sqrt{\Omega_\Lambda}]
\end{equation}
From Eqs.(\ref{eq:woleff3}) and (\ref{eq:olCT}), the equation of
state of dark energy is written as
\begin{equation}
\label{eq:wCT}
w_{\Lambda}^{eff}=-1+\frac{2}{3c}(1+z)\sqrt{\Omega_\Lambda}.
\end{equation}
The accelerating universe requires $w_{\Lambda}^{eff}<-1/3$, that is
$c>\sqrt{\Omega_\Lambda}(1+z)$. Combining Eqs. (\ref{eq:wmeff2}) and
(\ref{eq:olCT}), one obtains
\begin{equation}
\label{eq:wmCT}
w_{m}^{eff}=-\frac{2(1+z)\sqrt{\Omega_\Lambda}}{3c}\frac{\Omega_\Lambda}{1-\Omega_\Lambda}.
\end{equation}
Furthermore, one determine the Friedmann equation of this model via
Eqs. (\ref{eq:E2}), (\ref{eq:olCT}) and (\ref{eq:wmCT}).

\begin{figure}
 \centering
 \includegraphics[totalheight=3.5in, angle=0]{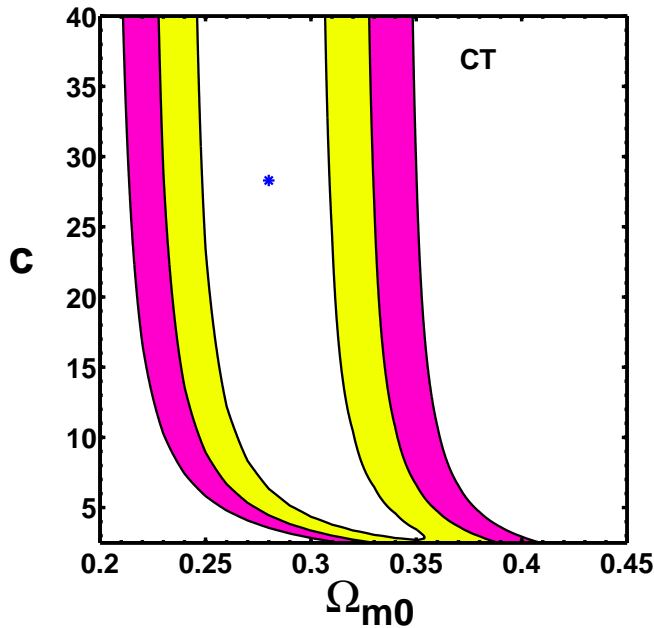}
 \caption{
Conformal time is taken as a cosmological time scale (CT for short).
The contours correspond to $68.3\%$, $95.4\%$ and $99.7\%$
confidence levels constrained from the recent observational data.
The blue dot marks the best-fit pair $(\Omega_{m0}, c)$. The fitting
values are $\Omega_{m0}=0.28^{+0.07}_{-0.03}$ and
$c=28.3^{+null}_{-25.5}$ ($68.3\%$ c.l.) with
$\chi^{2}_{min}=544.268$, where ``null'' denotes the absence of the
upper limit for $c$.}
 \label{fig:CT}
 \end{figure}

In Figure \ref{fig:CT}, we plot the probability contours constrained
from the recently compiled ``Union2 compilation'' of SNeIa data for
$68.3\%$, $95.4\%$ and $99.7\%$ confidence levels. The fitting
results are $\Omega_{m0}=0.28^{+0.07}_{-0.03}$ and
$c=28.3^{+null}_{-25.5}$ for $68.3\%$ confidence level with
$\chi^2_{min}=544.268$, where ``null'' denotes the absence of the
upper limit for $c$. We can see that the top sides of the three
contours also are open as the above model. Repeating the analysis
and calculations as done in the above case,
$\Delta\chi^2=\chi^2_{c=\infty}-\chi^2_{min}=0.151$ is also smaller
than $\Delta \chi^2_{1\sigma}(n=2)=2.30$. The further discussion
also will be shown in Section IV.

\section{Discussion and Conclusion}
In this letter, motivated by the fact that any nonzero $\Lambda$ can
introduce a length scale or a time scale into Einstein's theory,
$r_{\Lambda}=ct_{\Lambda}=\sqrt{3/|\Lambda|}$. Conversely, any
cosmological length scale or time scale can introduce a $\Lambda
(t)$, $\Lambda(t)=3/r^2_{\Lambda}(t)=3/(c^2t^2_{\Lambda}(t))$, we
investigate the $\Lambda (t)$CDM models corresponding to the length
scales, including the Hubble horizon, the particle horizon and the
future event horizon, and the time scales, including the age of the
universe and the conformal time. It comes out that an accelerating
expansion universe can be derived in  the cases taking the Hubble
horizon, the future event horizon, the age of the universe and the
conformal time as the length scale or time scale. Furthermore, the
modalities of the holographic dark energy and the agegraphic dark
energy can be derived from the $\Lambda (t)$CDM models. In this
scenario, the $\Lambda(t)$CDM model can be taken as the unified
origin of the holographic and agegraphic dark energy models with
interaction between the matter and the dark energy, where the
interacting term is determined by $Q=-\dot{\rho}_{\Lambda}$.

\begin{figure}
 \centering
 \includegraphics[totalheight=2.5in, angle=0]{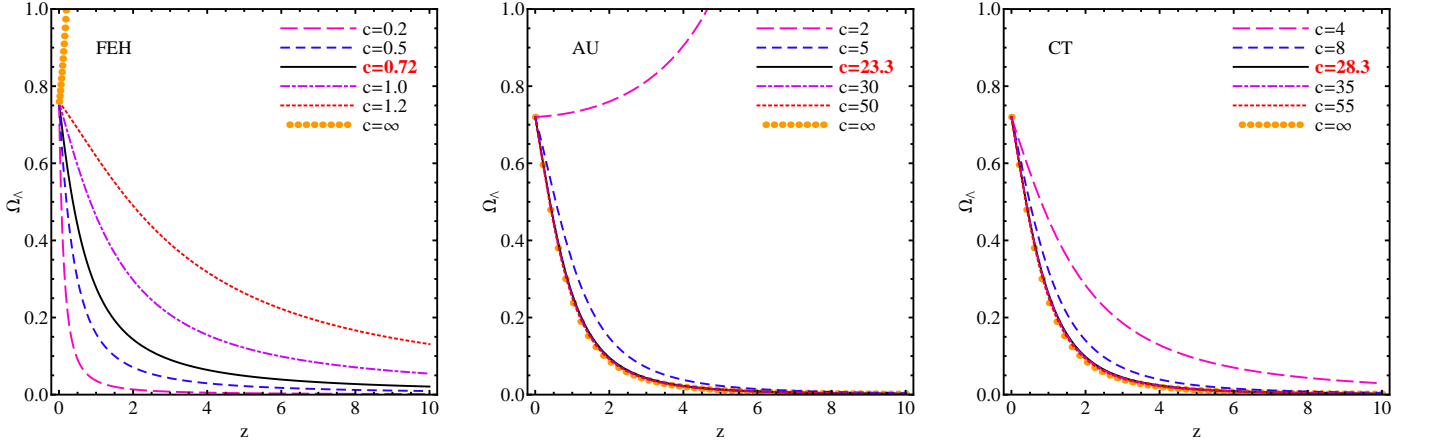}
 \caption{
The evolutions of $\Omega_{\Lambda}$ with respect to redshift $z$
with the variety of parameter $c$, corresponding to the future event
horizon as a cosmological length scale (FEH for short), the age of
the universe as a cosmological time scale (AU for short) and the
conformal time as a cosmological time scale (CT for short), where
$\Omega_{m0}$ takes the best-fit value for each case. The black
solid line in each panel is plotted with the best-fit pairs
$(\Omega_{m0},c)$.}
 \label{fig:ol_comb}
 \end{figure}

However, in the last two cases, the contours are not closed for
$68.3\%$, $95.4\%$ and $99.7\%$ confidence levels, which lead to the
absence of the upper limit for the parameter $c$. We found that this
problem chiefly originates from the evolution of $\Omega_{\Lambda}$
with respect to redshift $z$ with the variety of the $c$. In Figure
\ref{fig:ol_comb}, the evolutions of $\Omega_{\Lambda}$ with respect
to redshift $z$ with the variety of $c$ are displayed, corresponding
to the three cases taking the future event horizon, the age of the
universe and the conformal time as the length scale or time scale,
where $\Omega_{m0}$ takes the best-fit value for each case. The
black solid line in each panel is plotted with the best-fit pair
$(\Omega_{m0},c)$. In the first panel of Figure \ref{fig:ol_comb},
corresponding to take the future event horizon as the length scale,
the best-fit value $c=0.72$ lies in the range where
$\Omega_{\Lambda}$ is sensitive to the variety of $c$, which makes
the presence of the upper and lower limits for $c$ in $68.3\%$,
$95.4\%$ and $99.7\%$ confidence levels. In the second panel of
Figure \ref{fig:ol_comb}, corresponding to take the age of the
universe as the time scale, $\Omega_{\Lambda}$ is sensitive to the
variety of  $c$ when $c$ is small, however, $\Omega_{\Lambda}$ is
insensitive to the variety of $c$ when $c$ is big enough, further
more, the best-fit value $c=23.3$ lies in the range where
$\Omega_{\Lambda}$ is insensitive to the variety of $c$. As a
result, the lower limit of $c$ exits, but one cannot work out its
upper limit. In the third panel of Figure \ref{fig:ol_comb},
corresponding to take the conformal time as the time scale, the
situation is the same as the above one.

\begin{figure}
 \centering
 \includegraphics[totalheight=3.5in, angle=0]{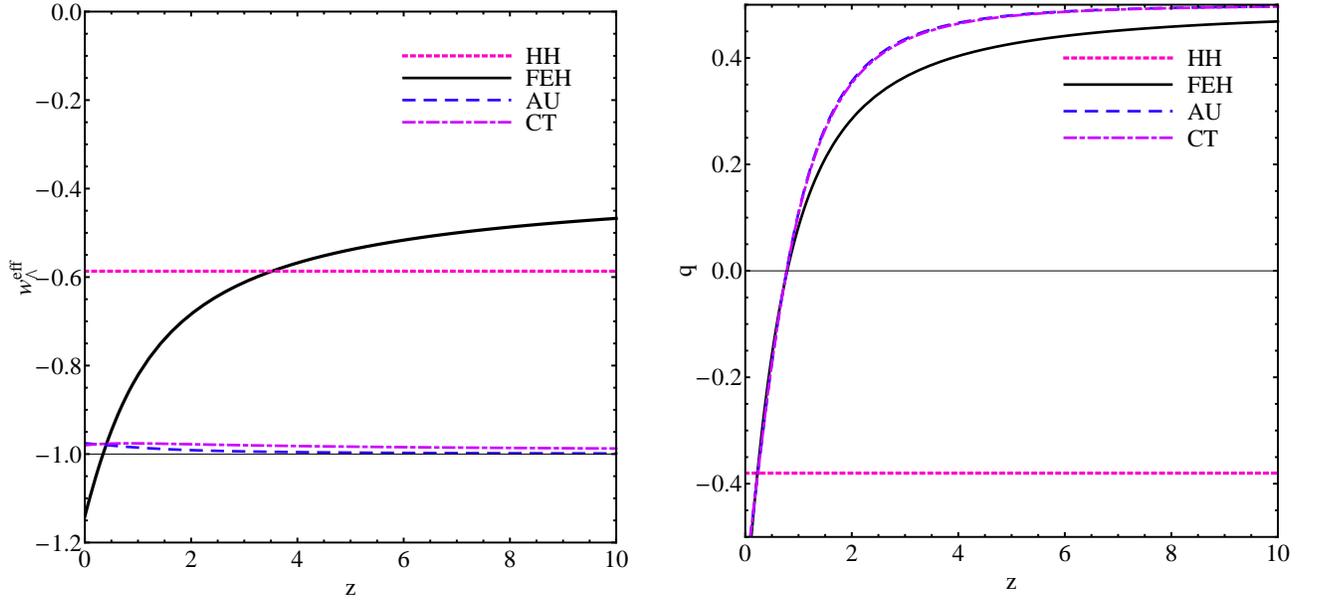}
 \caption{
The left panel displays the evolutions of $w_{\Lambda}^{eff}$ with
the corresponding best-fit $c$ or $(\Omega_{m0},c)$. The right panel
shows the evolutions of $q(z)$ with the corresponding best-fit $c$
or $(\Omega_{m0},c)$. HH, FEH, AU an CT are short for the Hubble
horizon, the future event horizon, the age of the universe and the
conformal time as a cosmological length scale or time scale,
respectively.}
 \label{fig:w_q_comb}
 \end{figure}

The left panel of Figure \ref{fig:w_q_comb} displays the evolutions
of $w_{\Lambda}^{eff}$ with the best-fit $c$ or $(\Omega_{m0},c)$,
corresponding to the four cases that can lead to the accelerating
expansion universe. In the case of taking the Hubble horizon as
length scale, $w_{\Lambda}^{eff}=-0.587^{+0.025}_{-0.023}$ is
quintessence like. $w_{\Lambda}^{eff}$ crosses the phantom divide in
the case of taking the future event horizon as a length scale. In
the other two cases, $w_{\Lambda}$ are big than $-1$ all through and
their variations with the variety of $z$ are not significant. The
right panel of Figure \ref{fig:w_q_comb} shows the evolutions of
$q(z)$ with the corresponding best-fit $c$ or $(\Omega_{m0},c)$ for
the four cases. In the case of taking the Hubble horizon as length
scale, $q=-0.380^{+0.037}_{-0.034}$ is a constant. A problem
deserves to be pointed out here. In this case, as it was discussed
in \cite{xu09c}, when $c$ is a fixed constant, non-transition from
decelerated expansion to accelerated expansion can be realized. The
author of \cite{xu09c} proposed a possible remedy, that is to
consider a time variable $c$, to solve this issue. Obviously, the
evolutions of $q(z)$ are very similar in the last three cases. As
$z$ grows, $q$ goes to $0.5$ at last, that is consistent with the
$\Lambda$CDM model.

 {\bf Acknowledgments.}
This work was supported by the National Natural Science Foundation
of China under the Distinguished Young Scholar Grant 10825313 and
Grant 11073005, and by the Ministry of Science and Technology
national basic science Program (Project 973) under grant No.
2007CB815401. Yun Chen would like to thank Bharat Ratra for his
kindly help, and to thank Xing Wu, Hongbao Zhang, Hongsheng Zhang
and Hao Wei for their helpful discussions and suggestions. LX was
supported by NSF (10703001) and SRFDP (20070141034) of P.R. China.
 JSA acknowledges financial support from CNPq-Brazil under grant Nos. 304569/2007-0 and 481784/2008-0.


\begin{thebibliography}{99}

\bibitem{SN}
A.~G.~Riess {\it et al.},
Astron.\ J.\  {\bf 116}, 1009 (1998),  [astro-ph/9805201];\\
S.~Perlmutter {\it et al.}, Astrophys.\ J.\  {\bf 517}, 565  (1999)
[astro-ph/9812133].

\bibitem{CMB}  D.~N.~Spergel {\it et al.}, Astrophys.\ J.\ Suppl.\ {\bf
148}, 175  (2003), [astro-ph/0302209].

\bibitem{BAO}  D.~J.~Eisenstein {\it et al.}, Astrophys.\ J.\ {\bf
633}, 560  (2005), [astro-ph/0501171].

\bibitem{Frieman08} J.~Frieman, M.~Turner and D.~Huterer, Ann.\ Rev.\ Astron.\ Astrophys.\ {\bf
46}, 385 (2008), [arXiv:0803.0982]

\bibitem{Tsujikawa2010} S.~Tsujikawa, [arXiv:1004.1493]

\bibitem{Rev}  E.~J.~Copeland, M.~Sami and S.~Tsujikawa, Int.\ J.\ Mod.\ Phys.\ D {\bf 15}, 1753 (2006), [hep-th/0603057];\\
R.~R.~Caldwell and M.~Kamionkowski, Ann. Rev. Nucl. Part. Sci. 59
(2009) 397, [arXiv:0903.0866].

\bibitem{ccp}  S.~Weinberg, Rev.\ Mod.\ Phys. {\bf 61}, 1 (1989);\\
 S.~Weinberg, [astro-ph/0005265];\\
 I.~Zlatev, L.~Wang and P.~J.~Steinhardt, Phys.\ Rev.\ Lett. {\bf 82}, 896  (1999), [astro-ph/9807002]

\bibitem{accp}
 S.~Weinberg, [astro-ph/0005265];\\
 A.~Vilenkin, [hep-th/0106083];\\
 J.~Garriga, M.~Livio and A.~Vilenkin, Phys.\ Rev. D {\bf 61}, 023503  (2000), [astro-ph/9906210];\\
 J.~Garriga and A.~Vilenkin, Phys.\ Rev.\ D {\bf 64}, 023517  (2001), [hep-th/0011262];\\
 L.~Amendola, Phys.\ Rev.\ D {\bf 62}, 043511  (2000), [astro-ph/9908023];\\
 G.~Caldera-Cabral, R.~Maartens and L.~A.~Urena-Lopez, Phys.\ Rev.\ D {\bf 79}, 063518  (2009), [arXiv:0812.1827];\\
 E.~J.~Copeland, M.~Sami and S.~Tsujikawa, Int.\ J.\ Mod.\ Phys.\ D {\bf 15},
 1753  (2006) , [hep-th/0603057].

 \bibitem{Peebles1988}P.~J.~E.~Peebles and B.~Ratra, \apj, {\bf
 325}, 17 (1988)

\bibitem{wDE-1}
M. $\ddot{\rm{O}}$zer and M. O. Taha, Phys. Lett. {\bf B171}, 363
(1986);\\
 M. $\ddot{\rm{O}}$zer and M. O. Taha, Nucl. Phys. B{\bf287}, 776 (1987);\\
 O. Bertolami, Nuovo Cimento Soc. Ital. Fis. {\bf B93}, 36 (1986);\\
K. Freese {\it et al}, Nucl. Phys.  {\bf B287}, 797 (1987);\\
M. S. Berman, Phys. Rev. {\bf43}, 1075 (1991);\\
J. C. Carvalho, J. A. S. Lima and I. Waga, Phys. Rev. {\bf{D46}}
2404 (1992);\\
 V.~Sahni and A.~Starobinsky, Int.J. Mod. Phys. D {\bf} 9 373 (2000)
 [astro-ph/9904398]

\bibitem{rhoP}  P.~Wang and X.-H. Meng, Class.Quant.Grav., {\bf 22}, 283 (2005), [astro-ph/0408495];\\
 H.~A.~Borges and S.~Carneiro, Gen. Rel. Grav. {\bf 37}, 1385
 (2005), [gr-qc/0503037];\\
  S.~Carneiro, M.~A.~Dantas, C.~Pigozzo and J.~S.~Alcaniz, Phys. Rev. D {\bf77}, 083504
  (2008), [arXiv:0711.2686];\\
  S.~Basilakos, M.~Plionis and J.~Sola, \prd, {\bf 80}, 083511
  (2009), [arXiv:0907.4555]

 \bibitem{Bousso08} R.~Bousso, Gen.\ Rel.\ Grav.\ {\bf 40}, 607  (2008), [arXiv:0708.4231].

 \bibitem{xu09a}L.~Xu, W.~Li,and J.~Lu, [arXiv:0905.4772].

 \bibitem{xu09b}L.~Xu, J.~Lu and W.~Li, Phys. Lett. B {\bf690}, 333 (2010), [arXiv:0905.4773].

\bibitem{Cohen99} A.~G.~Cohen,  D.~B.~Kaplan and  A.~E.~Nelson, Phys.\ Rev.\ Lett.\ {\bf
82}, 4971 (1999), [hep-th/9803132]

\bibitem{Li04} M.~Li, Phys.\ Lett.\ B {\bf 603}, 1  (2004), [hep-th/0403127]

\bibitem{holo}
 Q.-G.~Huang and Y.~Gong, JCAP {\bf 0408}, 006  (2004), [astro-ph/0403590];\\
 X.~Zhang and F.-Q.~Wu, Phys.\ Rev.\ D {\bf 72}, 043524 (2005), [astro-ph/0506310];\\
 Z.~Chang, F.-Q.~Wu and X.~Zhang, Phys.\ Lett.\ B {\bf 633}, 14 (2006), [astro-ph/0509531];\\
 X.~Zhang and F.-Q.~Wu, Phys.\ Rev.\ D {\bf 76}, 023502  (2007), [astro-ph/0701405];\\
 M.~Li {\it et al.}, JCAP, {\bf 0906}, 036  (2009), [arXiv:0904.0928];\\
M.~Li {\it et al.},  arXiv:0910.3855.

 \bibitem{Cai07} R.-G.~Cai, Phys.\ Lett.\ B, {\bf 657}, 228  (2007), arXiv:0707.4049.

 \bibitem{WZ07}H.~Wei and  S.~N.~Zhang, Phys.\ Rev.\ D, {\bf 76}, 063003 (2007), arXiv:0707.2129.

 \bibitem{WC08a}H.~Wei and R.-G.~Cai, Phys.\ Lett.\ B, {\bf 660}, 113 (2008), arXiv:0708.0884.

 \bibitem{agegra}
 X.~Wu {\it et al.}, [arXiv:0708.0349];\\
H.~Wei and R.-G.~Cai, Phys.\ Lett.\ B, {\bf 663}, 1 (2008),
arXiv:0708.1894.

 \bibitem{agegraP}

 K.~Y.~Kim, H.~W.~Lee and Y.~S.~Myung, Phys.\ Lett.\ B {\bf 660}, 118  (2008),arXiv:0709.2743;\\
K.~Y.~Kim {\it et al.}, Mod.\ Phys.\ Lett.\ A, {\bf 23}, 3049
(2008), arXiv:0803.0574.



\bibitem{Amanullah2010}  R.~Amanullah {\it et al.}, Astrophys.\ J.\ {\bf
716}, 712 (2010), arXiv:1004.1711.


\bibitem{MS09}Y.~S.~Myung and M.-G.~Seo, Phys.\ Lett.\ B, {\bf 671}, 435 (2009), arXiv:0803.2913.


\bibitem{xu09c}  L.~Xu, JCAP, {\bf0909}, 016 (2009),
arXiv:0907.1709.

\end{thebibliography}
\end{document}